\documentclass[11pt]{article}
\usepackage[english]{babel}
\usepackage{authblk}
\usepackage{etoolbox}
\usepackage[a4paper,top=2cm,bottom=2cm,left=2.5cm,right=2.5cm,marginparwidth=1.75cm]{geometry}
\usepackage{setspace}
\usepackage{amsmath}
\usepackage{graphicx}
\usepackage{float}
\usepackage[colorlinks=true, allcolors=blue]{hyperref}
\usepackage{multicol}
\usepackage{ragged2e} 
\usepackage{tcolorbox} 
\usepackage[dvipsnames]{xcolor} 
\usepackage[
    style=numeric,
    maxnames=1,   % show only the first author
    minnames=1,   % truncate author lists with "et al."
    sorting=none  % keep references in citation order
]{biblatex}

\AtBeginEnvironment{abstract}{\normalsize}

\makeatletter
\renewcommand\AB@affilsepx{ \quad } % no newline, just space
\makeatother

\title{Why the Northern Hemisphere Needs a 30–40 m Telescope and the Science at Stake: Key Targets of Opportunity on Gas and Ice Giants and their satellites}
\author[1]{Ricardo Hueso}
\author[2]{Leigh N. Fletcher}
\author[3]{Damya Souami}

\author[3]{Thierry Fouchet}
\author[4]{Tristan Guillot}
\author[5]{Olivier Mousis}
\author[6]{Patrick G. J. Irwin}
\author[2,7]{Michael Roman}

\author[1]{Arrate Antuñano}
\author[3]{Athena Coustenis}
\author[8,9]{Julia de León}
\author[3]{Sonia Fornasier}
\author[3]{Emmanuel Lellouch}
\author[10]{Alice Lucchetti}
\author[11]{Noemí Pinilla-Alonso}
\author[12]{Don Pollacco}
\author[1]{Agustín Sánchez-Lavega}
\author[13]{Daniel Toledo}

\affil[1]{Escuela de Ingeniería de Bilbao, Universidad del País Vasco UPV/EHU, Bilbao, Spain.}
\affil[2]{School of Physics and Astronomy, University of Leicester, Leicester, UK.}
\affil[3]{LIRA, Observatoire de Paris, CNRS, Université PSL, Meudon, France.}
\affil[4]{CNRS, Laboratoire Lagrange, Nice, France}
\affil[5]{Atmospheric, Oceanic and Planetary Physics, Department of Physics, University of Oxford, UK} 
\affil[6]{Solar System Science and Exploration Division, Southwest Research Institute, Boulder, CO, USA.}
\affil[7]{Facultad de Ingeniería y Ciencias, Universidad Adolfo Ibáñez, Santiago, Chile.}
\affil[8]{Instituto de Astrofísica de Canarias - IAC, Tenerife, Spain}
\affil[9]{Departamento de Astrofísica, Universidad de La Laguna, Tenerife, Spain}
\affil[10]{INAF-OAPD Astronomical Observatory of Padova, Padova, Italy}
\affil[11]{Instituto de Ciencias y Tecnologías Espaciales de Asturias (ICTEA)
Universidad de Oviedo, Spain}
\affil[12]{Department of Physics, University of Warwick, Coventry,UK}
\affil[13]{Instituto Nacional de Técnica Aeroespacial (INTA), Madrid, Spain.}

\date{}  % optional, remove date

\begin{document}
\maketitle

\begin{figure}[H]
\centering
\includegraphics[width=16.0cm]{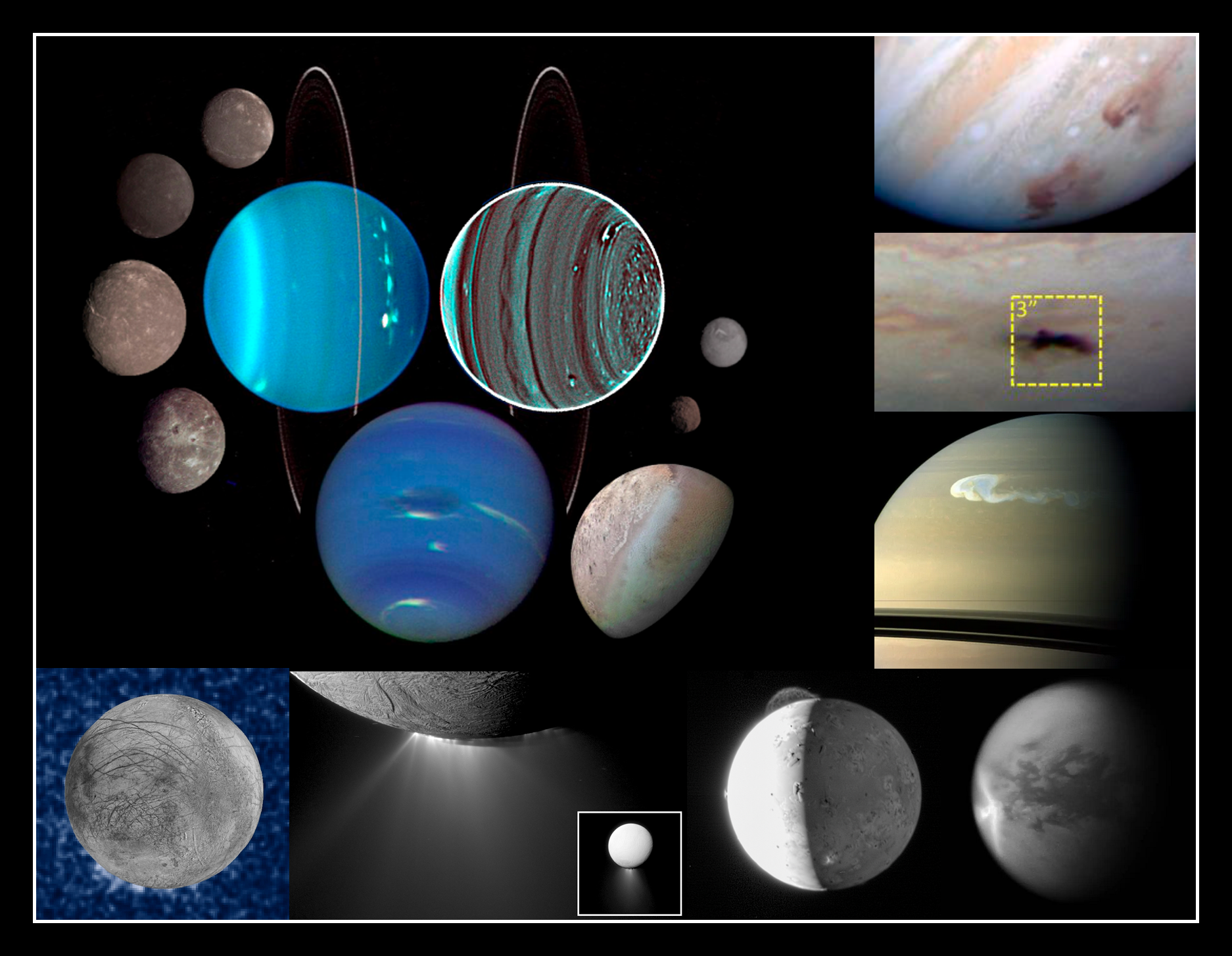}
%\caption{Icy giants, their satellite systems and unexpected target of interests including impacts in Jupiter, giant storms in Jupiter, Saturn and Titan, cryovolcanic activity in Europa an Enceladus and giant eruptions on Io.}
\end{figure}
\thispagestyle{empty}
\setcounter{figure}{0}  % Start next figure numbering from 1
% ---------------------------

\newpage

\setcounter{page}{1}

\begin{tcolorbox}[colback=RoyalBlue!5!white,colframe=black!75!black, 
    boxsep=1pt,          % overall inner padding
    left=6pt,            % optional fine tuning
    right=6pt,
    top=4pt,
    bottom=4pt,
    width=\textwidth]
\justifying
\noindent 
The Extremely Large Telescope (ELT) will transform our knowledge of the outer planets and their satellite systems; however the visibility of unique targets of opportunity with high scientific value will be reduced for northern objects. Uranus' declination favors observations from the Northern Hemisphere until 2055, and Neptune will be favored from the Northern Hemisphere from 2027 for the next 90 years. Jupiter and Saturn experience cycles of better observability from either hemisphere on cycles of 10 and 30 years. These planets and their satellite systems often offer unique opportunities for discovery through time-critical observations. \textbf{We argue that a 30-m class size telescope in the Northern Hemisphere with complementary scientific instrumentation to that on the ELT will secure the possibility of observing high-impact unpredictable phenomena in these systems.}
\end{tcolorbox}

\begin{multicols}{2}

\vspace{1mm}
\noindent\textbf{Solar System Science in the 2040s}
\vspace{1mm}

Breakthrough science in the Solar System in the 2040s will be accomplished with a combination of space missions, space telescopes, and extremely large ground-based telescopes using Adaptive Optics (AO) \cite{Fletcher2024, Kokotanekova2022}. 
Two key science themes in Solar System science in that decade will be: \textbf{(1)} The exploration of the moons of the giant planets and their potential for habitability and biosignatures \cite{ESA_L4_2024}, and \textbf{(2)} The in-depth characterization of Uranus and Neptune, which remain largely unexplored and hold fundamental information about the formation of the Solar System, being our closest and best examples of a class of planetary objects that dominates the census of exoplanets \cite{decadal}. 

%being also close examples in our planetary neighborhood of frequent exoplanets \cite{decadal}. 

While there are no space missions selected to Uranus and Neptune, missions to these targets are some of the highest priorities for space agencies \cite{decadal, NASA_UOP_2023}. Key areas of study for ice giants are their planetary origins \cite{Helled2020}, atmospheric dynamics \cite{Sanchez-Lavega2023}, and unique magnetospheres that interact in complex ways with the solar wind and upper atmosphere \cite{Stallard2025}. 
Exploring Uranus' moons and Neptune's captured dwarf planet Triton will provide crucial insights into the formation and activity of ocean worlds in our solar system. While Jupiter's moons will be explored by the JUpiter ICy moons Explorer (JUICE) and Europa Clipper missions in the 2030s \cite{Grasset2013, Pappalardo2024}, and ESA is developing a mission to Enceladus in the 2050s \cite{ESA_L4_2024}, their geological activity is time-dependent and is unconstrained by existing observations.

\vspace{1mm}
\noindent 
\textbf{Observability of Outer Planets and their moon systems}
\vspace{1mm}

\begin{figure}[H]
\centering
\includegraphics[width=6.2cm]{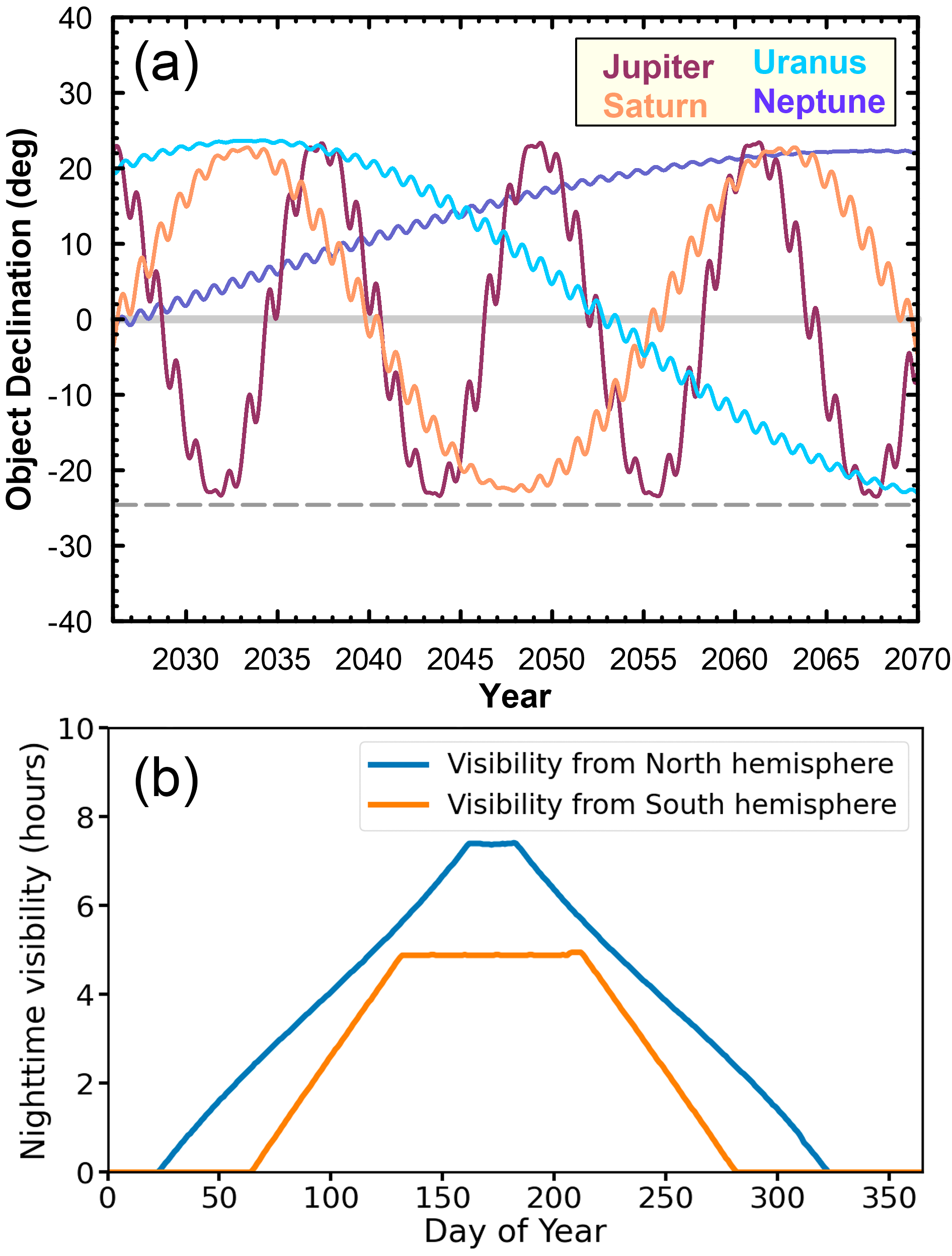}
\caption{(a) Declination of the outer planets and their satellite systems. %A gray line at 0 is shown for reference. 
The dashed gray line shows the latitude of Cerro Armarzones. (b) Number of nighttime hours for a target with declination 23.7$^{\circ}$ at elevations above 30$^\circ$ from equivalent latitudes at $\pm24.6^\circ$. Stronger differences can be found for targets with oppositions near the Northern Winter.}
\label{fig:declination_hours}
\end{figure}

Figure \ref{fig:declination_hours}a shows the declination of the Outer planets. Saturn, and thus Enceladus and Titan, are  favored from the south hemisphere in the 2040s, but the situation inverts in 2055. Uranus and Neptune are largely favored in the Northern Hemisphere over multiple decades, and Jupiter experiences periods of 5 yr of better observability from either hemisphere. AO systems are more efficient for high elevation targets ($>30^{\circ}$) \cite{Davies2012}. This has
a strong impact on the number of nights and number of hours per night each target is accessible for astronomical observations with AO (see \ref{fig:declination_hours}b).

\noindent \textbf{Critical Targets of Opportunity}
\vspace{1mm}

Unexpected solar system phenomena observed in the past have triggered target of opportunity observations (ToOs) that have led to important discoveries. Solar system ToOs typically require targeting specific locations on a rotating planet surface (e.g., giant impacts \cite{Harrington2004, Sanchez-Lavega2010} and convective storms \cite{Sanchez-Lavega2011}), or specific moments in time, such as giant eruptions on Io \cite{deKleer2014}, Io in eclipse from Jupiter, or minor satellites of Uranus and Neptune at their largest elongation from the planet. The following is a set of examples of rare phenomena that cannot be anticipated, require time-critical observations, and can provide high-impact advances to our understanding of planetary processes.

\vspace{2mm}
\noindent 
\textbf{Impacts on Giant Planets:} Impacts on giant planets are rare phenomena able to provide unique insights into impactor populations, the physics of high-velocity atmospheric impacts, and giant planet stratospheric chemistry and circulation. Two large impacts have been witnessed on Jupiter from comet SL9 \cite{Harrington2004} and an asteroid object in 2010 \cite{Sanchez-Lavega2010, Orton2011}. Statistics of these and smaller impacts \cite{Sanchez-Lavega2010, Hueso2018} imply that Jupiter may receive impacts from objects that can leave a visible trace in Jupiter's atmosphere once per decade. The observation of these events could significantly advance our knowledge of cosmic impacts. However, depending on the size of the impactor, its atmospheric trace could disappear in days to weeks or months, and the most interesting science is linked to observations acquired as quickly as possible. A high-resolution  imaging and spectroscopic investigation of the contamination in the atmosphere can reveal the impactor direction, the energy released, the dynamics of the upper atmosphere from the evolution of aerosols, and the atmospheric penetration and density of the object. Spectroscopy can reveal the presence of hydrocarbons, nitriles, and CO, and the water and oxygen content of the impactor and the dark debris in the atmosphere can reveal its origins and chemical make-up. Given their effective area, impacts on Saturn, Uranus, and Neptune could occur from less than once per decade for Saturn to once per century for Uranus and Neptune.

\vspace{2mm}
\noindent 
\textbf{Outer planet storms:} Superstorm eruptions on Jupiter and Saturn, and the sudden formation of meteorological systems on the calmer Uranus and Neptune are key to understanding heat transport within hydrogen-dominated atmospheres. Outstanding questions include the development of moist convection under inhibition from molecular weight stratification \cite{Guillot1995}, and the basic nature and depth of vortices in Uranus and Neptune \cite{Wong2018}. A 30-m size class telescope can resolve the detailed vertical and horizontal structure of these features, reveal the desiccation of the atmosphere produced by convective storms \cite{Li2015, Guillot2020} and determine the energetics of these phenomena from their effect on winds and cloud top altitudes \cite{Sanchez-Lavega2011}. For Uranus and Neptune, 30-m class size telescopes will achieve spatial resolutions approaching the best observations achieved by the unique {\it Voyager-2} flyby.

\vspace{2mm}
\noindent 
\textbf{Extreme eruptions on Io:} Io's intense volcanism  is driven by tidal heating  from its orbit in Laplace resonance with Europa and Ganymede. The details of how tidal heating produces eruptions remain unclear. Spacecraft data and monitoring with 6-8-m size telescopes show persistent but highly variable activity, including rare outbursts that can briefly double Io’s thermal emission. Early detection of these events can yield clues about Io’s interior and magma generation. Telescopes in the 30-m size class will provide much sharper spatial and spectral data, will be able to resolve multiple hot spots within single lava lakes, and track rapid changes occurring over minutes to days \cite{deKleer2017}. Time-critical observations following the discovery of a major eruption on Io will deepen our understanding of Io’s volcanism and support studies of cryovolcanism on ocean worlds such as Europa.

\vspace{2mm}
\noindent 
\textbf{Cryovolcanism: Europa, Enceladus, Triton and small ocean worlds:} Geyser-like plumes on Jupiter's moon Europa are a debated topic. Their presence will be investigated by the Europa Clipper and JUICE missions in the early 2030s \cite{Roth2025, Grasset2013}. If Europa is an active cryovolcanic world, the activity will be tenuous, variable, and only accessible for observations with extremely large apertures. Infrared/NIR spectroscopy can characterize water emissions and organics emitted from Europa's interior.
Among the Icy moons with active cryovolcanism, Triton remains the most difficult target for investigation. {\it Voyager-2} discovered active geysers \cite{Soderblom1990}, but with a size of 0.12 arcsec when observed from Earth, no current facility (not even JWST) can investigate geysers on Triton, and only stellar occultations can unveil part of the atmospheric evolution over a Triton year. Neptune and Triton will remain difficult to observe from the ELT due to Neptune's declination, but a 30-m class size telescope in the Northern hemisphere will allow to map compositional contrasts on the Uranian and Neptunian satellites with a spatial resolution comparable to what JWST can do for the Galilean satellites today.

\vspace{2mm}
\noindent 
\textbf{Rings and small moons:}
Planetary rings provide natural laboratories for disk processes and clues to the origin and evolution of planetary systems. Rings are active systems that produce short-lived ring arcs, spokes and orbital variations in some of the minor moons \cite{Tiscareno2018}. Advances in the understanding of the rings of Neptune and the interactions between the unstable complex systems of Uranus moons require time-domain observations that would be complex to obtain from the southern hemisphere for decades.

\vspace{2mm}
Finally, northern and southern large telescopes will enable complementary spectral and imaging capabilities, contributing to deeper insights into long-term processes than a single instrumental setup in a single large telescope. 

\end{multicols}

% -------------------------------------------------------
% REFERENCES BLOCK
\begin{multicols}{1}
\setlength{\itemsep}{0pt}
% Print the bibliography
\printbibliography
%\bibliographystyle{abbrv}
%\bibliography{Bibliography.bib}
%\addbibresource{Bibliography.bib}
%\printbibliography
\end{multicols}
% -------------------------------------------------------

\end{document}